\documentclass[twocolumn,fp]{jpsj3}
\usepackage{graphicx}
\usepackage{dcolumn}
\usepackage{bm}
\usepackage{ulem}
\usepackage{braket}
\usepackage{amsmath}
\usepackage{color}

\title{Nonreciprocal Transport in Noncoplanar Magnetic Systems without Spin-Orbit Coupling, Net Scalar Chirality, or Magnetization}

\author{Satoru Hayami and Megumi Yatsushiro}
\inst{
Faculty of Science, Hokkaido University, Sapporo 060-0810, Japan 
}

\abst{
We propose a new mechanism of nonlinear nonreciprocal transport in magnetic systems. 
By considering a noncoplanar magnetic ordering on a bilayer triangular lattice, we clarify that a local scalar chirality degree of freedom is a source of nonreciprocal transport, which does not require any relativistic spin-orbit coupling, net scalar chirality, and net magnetization. 
We show a close relationship between the asymmetric band modulation and the nonreciprocal transport under the noncoplanar magnetic ordering based on the model-parameter dependences in the real-space picture. 
}

\begin{document}
\maketitle

\section{Introduction}

A noncoplanar spin texture has recently been attracted in condensed matter physics.  
Although its appearance in materials has been still rare compared to collinear and coplanar spin textures, the recent developments of experimental techniques push their searches and discoveries in a variety of materials, such as the all-in/all-out spin texture in the pyrochlore structure~\cite{Yamaura_PhysRevLett.108.247205,Sagayama_PhysRevB.87.100403,Disseler_PhysRevB.89.140413,Huang_PhysRevLett.112.167203,Donnerer_PhysRevLett.117.037201} and other cubic structures~\cite{Jensen_PhysRevB.23.6180,Kawarazaki_PhysRevLett.61.471,Shindou_PhysRevLett.87.116801,Alonso_PhysRevB.64.054408,Hayami_PhysRevB.89.085124} and the skyrmion crystal spin texture in both noncentrosymmetric and centrosymmetric crystal structures~\cite{Muhlbauer_2009skyrmion,yu2011near,seki2012observation,kurumaji2019skyrmion,hirschberger2019skyrmion,khanh2020nanometric,Yasui2020imaging,Tokura_doi:10.1021/acs.chemrev.0c00297,Hirschberger_10.1088/1367-2630/abdef9,khanh2022zoology,takagi2022square}.
Simultaneously, it has been revealed that these noncoplanar spin configurations are brought about by the synergy among magnetic interactions that originates from the internal electronic degrees of freedom, such as frustrated exchange interaction and charge-spin-orbital coupled interaction in real and momentum spaces~\cite{rossler2006spontaneous,Binz_PhysRevB.74.214408,Gardner_RevModPhys.82.53,lacroix2011introduction,nagaosa2013topological,Akagi_PhysRevLett.108.096401,Hayami_PhysRevB.95.224424,batista2016frustration,hayami2021topological,yambe2022effective}. 

Noncoplanar magnetic orderings manifest themselves not only in unusual magnetism but also in anomalous transport properties. 
The most typical example is the topological Hall effect, where conduction electrons acquire a gauge flux through the spin Berry phase under the noncoplanar magnetic orderings~\cite{Loss_PhysRevB.45.13544,Ye_PhysRevLett.83.3737,Ohgushi_PhysRevB.62.R6065,taguchi2001spin,tatara2002chirality,machida2007unconventional,Neubauer_PhysRevLett.102.186602,takatsu2010unconventional,ueland2012controllable,Hamamoto_PhysRevB.92.115417,nakazawa2018topological,Matsui_PhysRevB.104.174432}. 
There, the key ingredient is scalar spin chirality (SSC) defined as the triple scalar product of spins as $\bm{s}_i \cdot (\bm{s}_j \times \bm{s}_k)$ for spin $\bm{s}_i$ at site $i$. 
From the symmetry aspect, the emergence of the topological Hall effect under nonzero SSC is natural, since its net component has the same symmetry as a time-reversal-odd axial-vector quantity like the uniform magnetization. 
In other words, the topological Hall effect occurs in the absence of the net magnetization once the net SSC becomes nonzero in the system, as demonstrated by various noncoplanar orderings in different lattice systems, such as the face-centered-cubic~\cite{Shindou_PhysRevLett.87.116801,Hayami_PhysRevB.90.060402}, triangular~\cite{Martin_PhysRevLett.101.156402,Akagi_JPSJ.79.083711,Kato_PhysRevLett.105.266405,Akagi_PhysRevLett.108.096401,hayami_PhysRevB.91.075104,Hayami_PhysRevB.94.024424,Ozawa_PhysRevLett.118.147205}, honeycomb~\cite{Venderbos_PhysRevB.93.115108}, Shastry-Sutherland~\cite{shahzad2017phase}, kagome~\cite{Barros_PhysRevB.90.245119,Ghosh_PhysRevB.93.024401}, and square  lattices~\cite{Wang_PhysRevB.103.104408,hayami2022multiple}. 

In the present study, we investigate another transport phenomenon under the noncoplanar magnetic orderings with the SSC degree of freedom by focusing on nonlinear nonreciprocal transport against the second-order electric field~\cite{rikken1997observation, Rikken_PhysRevLett.87.236602,tokura2018nonreciprocal}. 
In contrast to the linear Hall effect, the nonreciprocal transport is caused when the spatial inversion ($\mathcal{P}$) symmetry is broken in addition to the time-reversal ($\mathcal{T}$) symmetry~\cite{togawa2016symmetry,Holder_PhysRevResearch.2.033100,Fei_PhysRevB.102.035440,Watanabe_PhysRevResearch.2.043081,Watanabe_PhysRevX.11.011001,Yatsushiro_PhysRevB.104.054412}. 
For example, swirling spin textures with the magnetic toroidal moment as polar-vector quantity accompanied by spirals and magnetic skyrmions are prototypes to exhibit nonreciprocal transport~\cite{Seki_PhysRevB.93.235131,giordano2016spin,yokouchi2018current,Hoshino_PhysRevB.97.024413,Aoki_PhysRevLett.122.057206,jiang2020electric,seki2020propagation,ishizuka2020anomalous,Akaike_PhysRevB.103.184428,bhowal2022magnetoelectric,hayami2022helicity}. 
The emergence of the nonreciprocal transport is microscopically understood from an asymmetric modulation of the electronic and magnon band structures in the presence of the relativistic spin-orbit coupling (SOC)~\cite{Yanase_JPSJ.83.014703, Hayami_PhysRevB.90.024432,Hayami_doi:10.7566/JPSJ.84.064717, Iguchi_PhysRevB.92.184419, Hayami_doi:10.7566/JPSJ.85.053705, Gitgeatpong_PhysRevLett.119.047201,sato2019nonreciprocal,yatsushiro2019atomic,Matsumoto_PhysRevB.101.224419, Matsumoto_PhysRevB.104.134420, Hayami_PhysRevB.105.014404,yatsushiro2021microscopic}. 
Meanwhile, it was recently shown that the asymmetric band structure can be engineered by the SOC-free antiferromagnets (AFMs) with the aid of the local SSC~\cite{Hayami_PhysRevB.101.220403, Hayami_PhysRevB.102.144441, Hayami_PhysRevResearch.3.043158,hayami2021phase, Hayami_PhysRevB.105.024413}, which results in the nonreciprocal transport in frustrated magnets under an external magnetic field even in the materials with the negligibly small SOC~\cite{comment_Hayami}. 

To further gain an understanding of the relationship between nonreciprocal transport and local SSC, we construct a minimal and fundamental magnetic system, i.e., a noncoplanar magnetic ordering with the all-out-type spin configuration on a bilayer triangular lattice. 
We show that nonreciprocal transport emerges once such a magnetic ordering occurs. 
The mechanism lies in effective hopping arising from the local SSC, which does not require any SOC, net SSC, and even magnetization. 
We show the essential real-space hopping processes contributing to the asymmetric band structure and nonreciprocal transport. 
Our result provides a new transport function based on the SSC degree of freedom, which will be a useful reference to search further AFM materials applicable to spintronic devices.

\begin{figure}[ht!]
\begin{center}
\includegraphics[width=1.0 \hsize ]{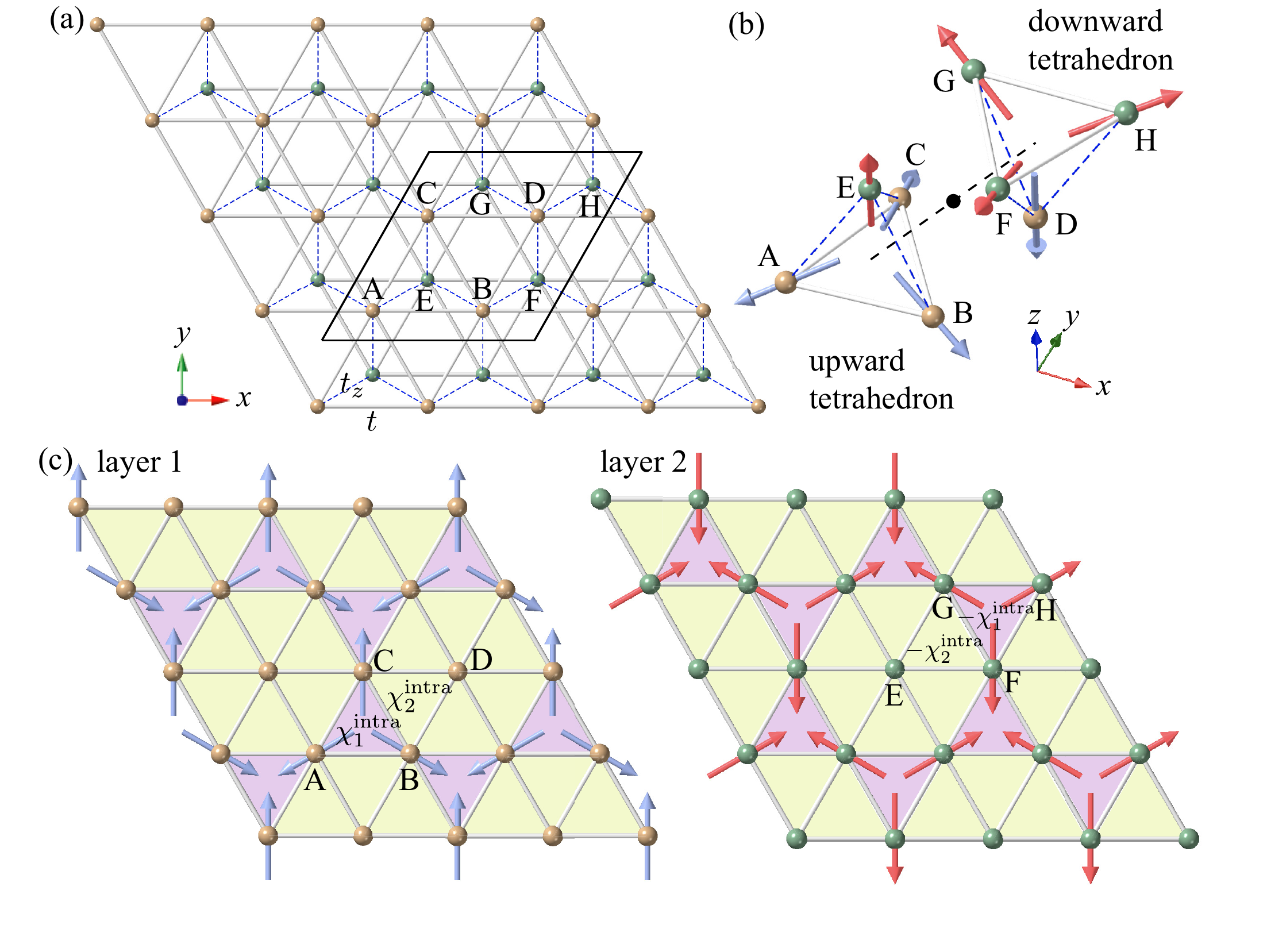} 
\caption{
\label{fig: fig1}
(Color online) 
(a) Bilayer triangular-lattice structure. 
The solid (dashed) bonds represent the intralayer (interlayer) hopping paths. 
The black solid parallelogram stands for the magnetic unit cell, which includes eight sublattices A--H. 
(b) Eight-sublattice spin texture. 
The black circle represents the inversion center of the crystal structure. 
The red (blue) arrow stands for the spin with the $+s^z$ ($-s^z$) component.
(c) The real-space inplane spin texture for layer 1 (left panel) and layer 2 (right panel), where $\chi^{\rm intra}_1$ and $\chi^{\rm intra}_2$ represent the local SSC in each triangle plaquette. 
}
\end{center}
\end{figure}

The rest of the paper is structured as follows. 
In Sect.~\ref{sec: Model}, we introduce a bilayer-triangular lattice model. 
We show that an eight-sublattice noncoplanar spin texture on the bilayer-triangular lattice is a minimum model to exhibit the nonreciprocal transport without SOC, net SSC, or magnetization. 
Then, we discuss the asymmetric band structure and nonreciprocal transport under the noncoplanar ordering in Sect.~\ref{sec: Results}. 
We find that the local SSC plays an important role in inducing SOC-free nonreciprocal transport. 
We summarize our results in Sect.~\ref{sec: Summary}.
In Appendices~\ref{sec: Model parameter dependence of nonreciprocal transport} and \ref{sec: Filling dependence of nonreciprocal transport}, we present the detailed model parameter dependences of the nonlinear conductivity in addition to the local SSC.

\section{Model}
\label{sec: Model}

We aim at investigating the relationship between nonlinear transport and local SSC. 
For that purpose, we suppose the situation where there are no net SSC and magnetization but local SSC under the magnetic orderings. 
Such a situation can be theoretically engineered by considering the two types of triangles consisting of three spins so as to have the local SSC with the opposite sign and by aligning them to vanish a net component. 
As an example, we consider a bilayer triangular-lattice system in Fig.~\ref{fig: fig1}, where the upper-layer sites (E, F, G, and H) are located at the center of the upward triangle in the lower layer consisting of A, B, C, and D. 
We hereafter refer to the lower (upper) layer as layer 1 (layer 2). 
We take the length between the nearest-neighbor sites in the same plane as the length unit and also set the length between E and the center of the upward triangle ABC as the unity. 
The tight-binding Hamiltonian is given by 
\begin{align}
\mathcal{H}= - \sum_{ij\sigma}t_{ij} c_{i\sigma}^{\dagger}c_{j\sigma}^{} - \sum_{i\sigma \sigma'} \bm{h}_i \cdot  c_{i\sigma}^{\dagger} \bm{\sigma}_{\sigma \sigma'}c_{i\sigma'}^{},
\label{eq: Ham}
\end{align}
where $c^{\dagger}_{i\sigma}$ ($c_{i\sigma}^{}$) is the creation (annihilation) operator for site $i$ and spin $\sigma=\uparrow, \downarrow$. 
The first term in Eq.~(\ref{eq: Ham}) represents the kinetic energy of electrons; we consider the nearest-neighbor hopping $t$ ($t_z$) in the intralayer (interlayer). 
The second term in Eq.~(\ref{eq: Ham}) stands for the site-dependent mean-field term $\bm{h}_i$ that arises from the Coulomb interaction; $\bm{\sigma}$ is the vector of the Pauli matrices. 
To investigate the situation where no net SSC and magnetization are present, we consider an eight-sublattice AFM ordering consisting of A--H in Fig.~\ref{fig: fig1}(a), whose spin texture is characterized by the all-out-type one in Fig.~\ref{fig: fig1}(b): $\bm{h}_{\rm A}= h(-\sqrt{2/3}, -\sqrt{2}/3,-1/3)$, $\bm{h}_{\rm B}= h(\sqrt{2/3}, -\sqrt{2}/3,-1/3)$, $\bm{h}_{\rm C}= h(0, 2\sqrt{2}/3,-1/3)$, $\bm{h}_{\rm D}= h(0, 0,-1)$, $\bm{h}_{\rm E}= -\bm{h}_{\rm D}$, $\bm{h}_{\rm F}= -\bm{h}_{\rm C}$, $\bm{h}_{\rm G}= -\bm{h}_{\rm B}$, and $\bm{h}_{\rm H}= -\bm{h}_{\rm A}$. 
Although there is an inversion center between the upward tetrahedron ABCE and downward tetrahedron DFGH at $\bm{h}_i=\bm{0}$ as shown by the black circle in Fig.~\ref{fig: fig1}(b), the onset of the magnetic ordering leads to the breaking of the $\mathcal{P}$ symmetry while keeping the $\mathcal{PT}$ symmetry. 
It is noted that the site symmetry at sites D and E is different from the others under this noncoplanar ordering; the expectation values of spins at sites D and E are not equal to the others, i.e., $|\langle \bm{s}_{\rm A} \rangle|=|\langle \bm{s}_{\rm B} \rangle|=|\langle \bm{s}_{\rm C} \rangle|=|\langle \bm{s}_{\rm F} \rangle|=|\langle \bm{s}_{\rm G} \rangle|=|\langle \bm{s}_{\rm H} \rangle|$, $|\langle \bm{s}_{\rm D} \rangle|=|\langle \bm{s}_{\rm E} \rangle|$, and $|\langle \bm{s}_{\rm A} \rangle|\neq |\langle \bm{s}_{\rm D} \rangle|$ where $\bm{s}_{i} = (1/2)\sum_{\sigma\sigma'}c_{i\sigma}^{\dagger} \bm{\sigma}_{\sigma \sigma'}c_{i\sigma'}^{}$. 
In addition, the system does not show a net magnetization; $\sum_i \langle \bm{s}_i \rangle = \bm{0}$. 
In the following, we set $t=1$ as the energy unit of the Hamiltonian in Eq.~(\ref{eq: Ham}) and fix $t_z=0.8$.

The $\mathcal{P}$-symmetry breaking is also found in the real-space SSC distribution, which is calculated by the triple scalar product of spin moments, $\chi_{ijk}=\langle \bm{s}_{i}\rangle \cdot (\langle \bm{s}_{j}\rangle \times \langle \bm{s}_{k}\rangle)$ in each plaquette for layer 1 (layer 2) in the left (right) panel of Fig.~\ref{fig: fig1}(c). 
In each layer, there are two types of triangles to have different SSCs with $|\chi^{\rm intra}_{1}|$ (pink triangle) and $|\chi^{\rm intra}_2|$ (yellow triangle), whose signs are opposite in the two layers, as shown in Fig.~\ref{fig: fig1}(c).  
As the $\mathcal{P}$ operation that transforms between (A, B, C, D) and (H, G, F, E) does not change the sign of $\chi^{\rm intra}_{1}$ and $\chi^{\rm intra}_{2}$, the SSC distribution has odd parity regarding the $\mathcal{P}$ operation. 
Meanwhile, the combination of $\mathcal{P}$ and $\mathcal{T}$ operations recovers the original SSC distribution, since the sign of the SSC is reversed under the $\mathcal{T}$ operation.

Although we here focus on the transport property by supposing the eight-sublattice noncoplanar ordering in Figs.~\ref{fig: fig1}(a) and \ref{fig: fig1}(b), we discuss two possible scenarios to stabilize this noncoplanar ordering. 
One scenario is the multiple-$Q$ instability. 
From the momentum-space viewpoint, the noncoplanar ordering is characterized by the triple-$Q$ state with equal intensities at three wave vectors, $\bm{k}=\bm{b}_1/2$, $\bm{b}_2/2$, and $-(\bm{b}_1+\bm{b}_2)/2$ where $\bm{b}_\lambda$ ($\lambda=1$--$3$) are the reciprocal vectors, in addition to the uniform component $\bm{k}=(0,0)$.  
Since a similar triple-$Q$ spin configuration, which is obtained by reversing the directions of $\bm{s}_{\rm D}$ and $\bm{s}_{\rm E}$ so as to vanish the intensity at $\bm{k}=(0,0)$, is stabilized when considering the effect of the ring-exchange interaction~\cite{Momoi_PhysRevLett.79.2081} and Fermi surface instability~\cite{Martin_PhysRevLett.101.156402,Akagi_JPSJ.79.083711,Kato_PhysRevLett.105.266405,Kumar_PhysRevLett.105.216405,Akagi_PhysRevLett.108.096401,hayami_PhysRevB.91.075104,Hayami_PhysRevB.94.024424}, the present triple-$Q$ spin configuraion might be stabilized by taking into account additional exchange interaction and/or anisotropy to induce the $\bm{k}=(0,0)$ component. 
Another scenario is the symmetry lowering of the lattice structure in the paramagnetic state. 
For example, when we consider the lattice structure consisting of two tetrahedrons ABCE and DFGH by differently setting the intratetrahedron hopping and intertetrahedron hopping, the all-out spin texture in each tetrahedron is realized by considering the strong intratetrahedron AFM exchange interaction. 
In such a situation, the following result for the nonreciprocal transport remains the same qualitatively.

\begin{figure}[t!]
\begin{center}
\includegraphics[width=1.0 \hsize ]{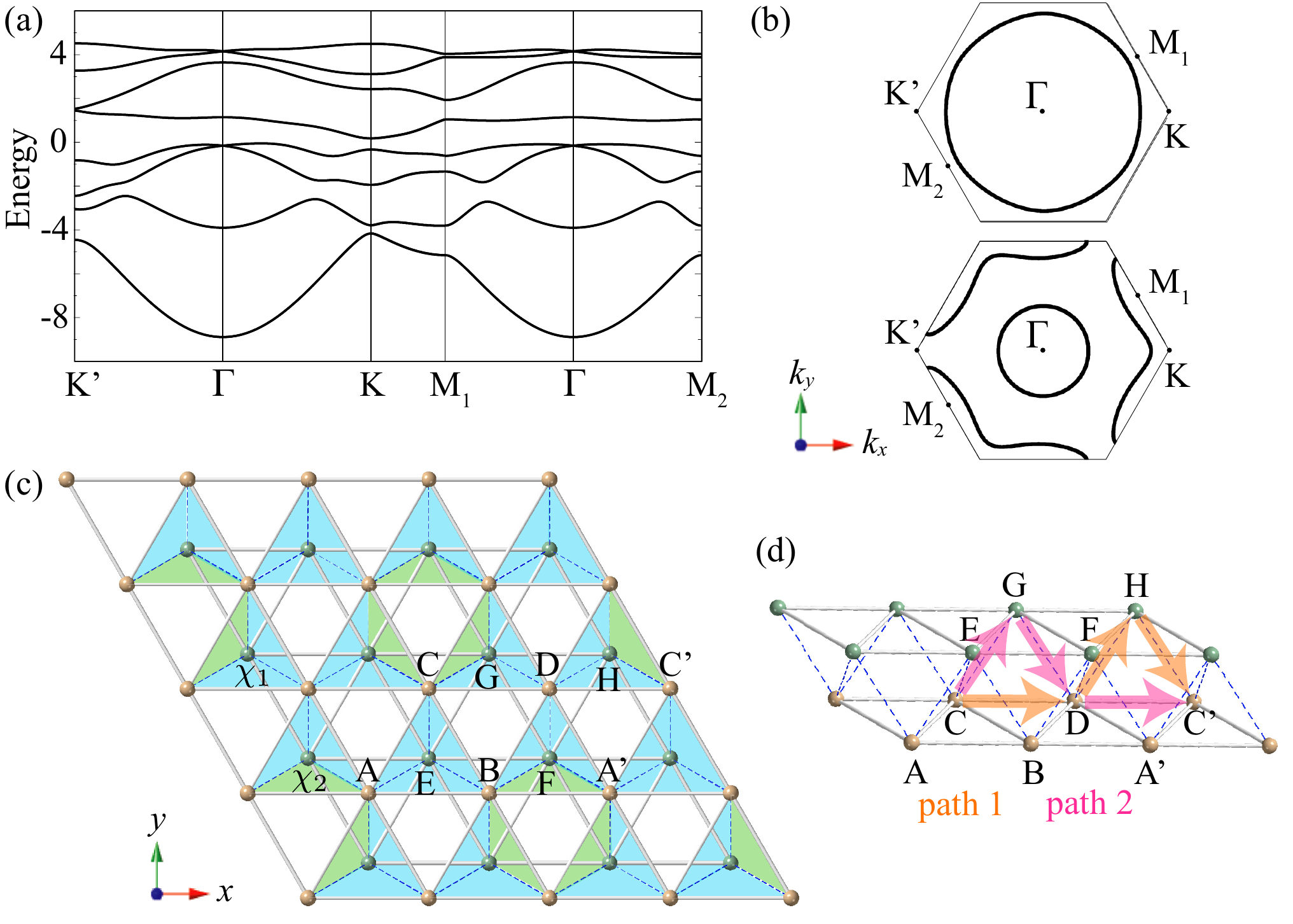} 
\caption{
\label{fig: fig2}
(Color online) 
(a) The band structure under the noncoplanar ordering in Fig.~\ref{fig: fig1}(b) at $h=2$. 
(b) The isoenergy surfaces in the magnetic Brillouin zone (a) at $n_{\rm e}=0.18$ (upper panel) and $n_{\rm e}=0.34$ (lower panel).
(c) SSC distribution [$\chi_1$ (blue) and $\chi_2$ (green)] across two triangular-lattice planes, where the spin texture is omitted. 
(d) Real-space two hopping processes (path 1 and path 2) contributing to the asymmetric band modulation along the $x$ direction. 
}
\end{center}
\end{figure}

\section{Results}
\label{sec: Results}

\subsection{Asymmetric band structure}
\label{sec: Asymmetric band structure}
Figure~\ref{fig: fig2}(a) shows the band structure under the noncoplanar ordering for $h=2$ along the high-symmetric lines of the magnetic Brillouin zone. 
Each band is doubly degenerate owing to the $\mathcal{PT}$ symmetry. 
One clearly finds that the asymmetric band modulation in terms of the $\Gamma$ point appears along the K'-$\Gamma$-K line, while the band is symmetric along the M$_1$-$\Gamma$-M$_2$ line. 
The angle dependence of the asymmetric band modulation is given by $k_x (k_x^2-3k_y^2)$, which satisfies the threefold rotational symmetry, as shown by the isoenergy surfaces in the cases of $n_{\rm e}=0.18$ and $n_{\rm e}=0.34$ in the upper and lower panels of Fig.~\ref{fig: fig2}(b), respectively. 
It is noted that there is a slight threefold band modulation at low filling $n_{\rm e}=0.18$. 
The emergence of the asymmetric band deformation corresponds to the active magnetic toroidal octupole $T_{3a}$~\cite{Hayami_PhysRevB.98.165110}. 

To examine the important model parameters for the asymmetric band deformation, we evaluate a quantity consisting of products among the $l$th power of the Hamiltonian matrix at wave vector $\bm{k}$, $\mathcal{H}^l(\bm{k})$, as ${\rm Tr}[\mathcal{H}^l (\bm{k})]-{\rm Tr}[\mathcal{H}^l (-\bm{k})]$~\cite{Hayami_PhysRevB.101.220403, Hayami_PhysRevB.102.144441,Hayami_PhysRevB.105.014404}. 
As a result, the lowest-order contribution is obtained as $h^3 t t_z^2 (\cos k_x - \cos \sqrt{3} k_y) \sin k_x$ in the order of $l=6$. 
The $h^3$ dependence indicates the necessity of the noncoplanar spin configuration, as detailed below. 
Moreover, one notices that the expression includes the interlayer hopping proportional to $t_z^2$ as well as the intralayer hopping $t$. 
In short, the local SSC extending over two layers connected by $t_z$ is essential to induce the asymmetric band modulation. 
Indeed, the local SSC defined on the interlayer triangle plaquette is distributed in a threefold symmetric way so that the inversion symmetry is broken, as shown in Fig.~\ref{fig: fig2}(c). 
We plot the local SSC defined by $\langle \bm{s}_{i}\rangle \cdot (\langle \bm{s}_{j}\rangle \times \langle \bm{s}_{k}\rangle)$ for $i,j \in$ A, B, C, D and $k \in$ E, F, G, H in Fig.~\ref{fig: fig2}(c), which are denoted by $\chi_1$ and $\chi_2$; $\chi_1$ consists of the triangles including site D or E, while $\chi_2$ does not contain both sites D and E. 
The SSC distribution for the other interlayer triangle plaquettes for $i,j \in$ E, F, G, H and $k \in$ A, B, C, D is obtained by reversing the sign of $\chi_1$ and $\chi_2$ owing to the $\mathcal{PT}$ symmetry.

To further investigate the dominant microscopic hopping processes, we explicitly write down the subscript for site index as $t \to t_{ij}$, $t_z \to t_{zij}$, and $\bm{h} \to \bm{h}_i$. 
By considering the hopping processes within four triangles (ABE, BA'F, CDG, and DC'H) along the $x$ direction in Fig.~\ref{fig: fig2}(c), one obtains the lowest-order contribution as $ h^y_{\rm C} h^z_{\rm D} (h^x_{\rm G}t_{z {\rm CG}} t_{z {\rm DG}}  t_{\rm C'D} +    h^x_{\rm H} t_{\rm CD} t_{z {\rm C'H}} t_{z {\rm DH}}) \sin 2 k_x$, whose schematic real-space processes are denoted by path 1 and path 2 in Fig.~\ref{fig: fig2}(d) (The contributions for the other directions are obtained by the threefold rotation operation). 
This expression indicates the importance of the local SSC, $\chi_1 = \langle \bm{s}_{{\rm C}}\rangle \cdot (\langle \bm{s}_{{\rm G}}\rangle \times \langle \bm{s}_{\rm D}\rangle)$ and $ \langle \bm{s}_{{\rm C}}\rangle \cdot (\langle \bm{s}_{{\rm D}}\rangle \times \langle \bm{s}_{\rm H}\rangle)$, which are proportional to $h^y_{\rm C} h^x_{\rm G} h^z_{\rm D} $ and $h^y_{\rm C} h^z_{\rm D} h^x_{\rm H}$, respectively, as well as the interlayer hopping paths $t_{z {\rm CG}} t_{z {\rm DG}}$ and $t_{z {\rm C'H}} t_{z {\rm DH}}$. 
Meanwhile, in the lowest order, the hopping processes within two triangles (ABE and BFA'), which include $\chi_2$, do not contribute to the asymmetric band deformation. 

\begin{figure}[t!]
\begin{center}
\includegraphics[width=1.0 \hsize ]{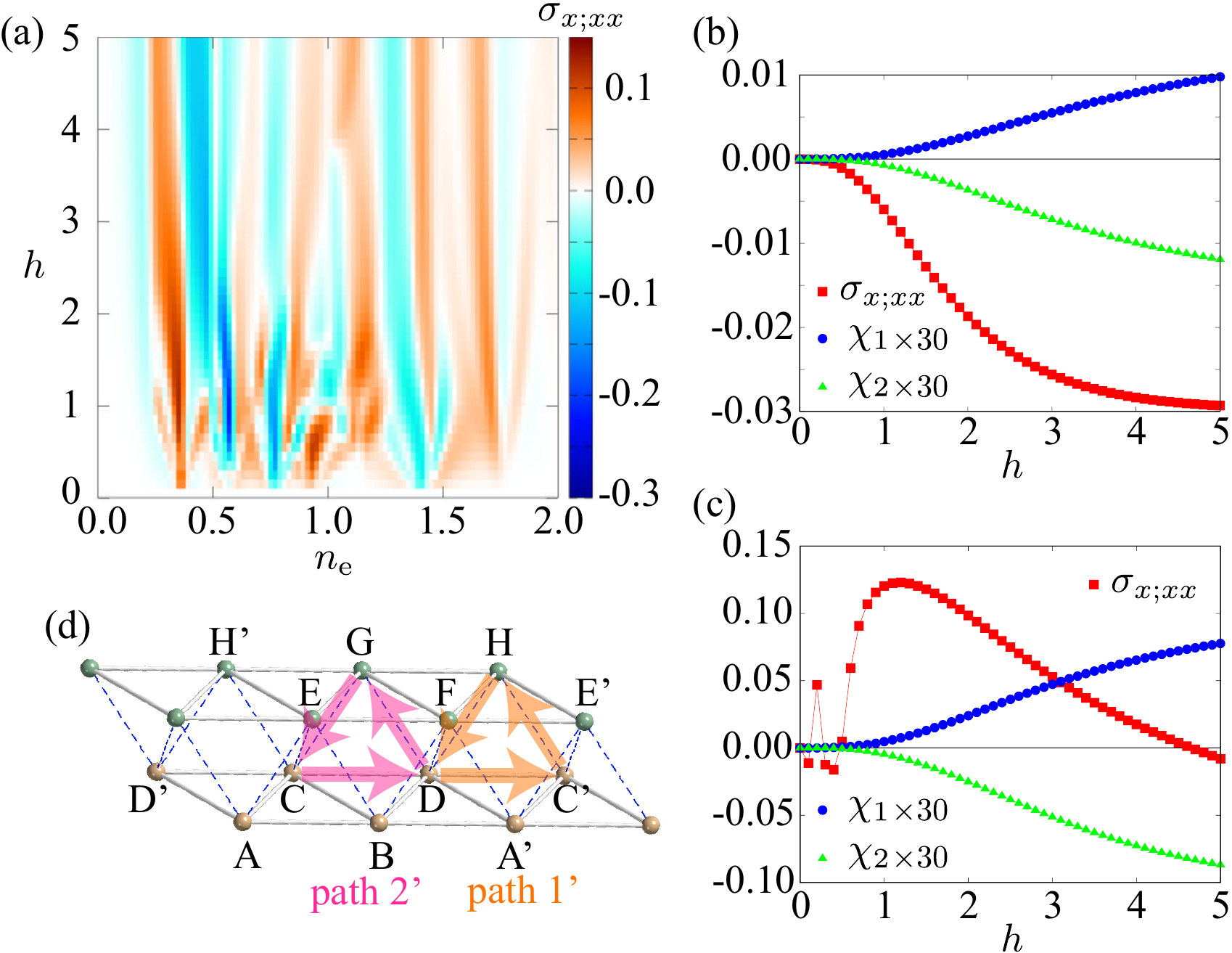} 
\caption{
\label{fig: fig3}
(Color online) 
(a) Contour plot of $\sigma_{x;xx}$ in the plane of $n_{\rm e}$ and $h$. 
(b,c) $h$ dependences of $\sigma_{x;xx}$, $\chi_1$, and $\chi_2$ for (b) $n_{\rm e}=0.18$ and (c) $n_{\rm e}=0.34$. 
(d) Two hopping processes (path 1' and path 2') contributing to $\sigma_{x;xx}$. 
}
\end{center}
\end{figure}

\subsection{Nonreciprocal transport}
\label{sec: Nonreciprocal transport}

Next, we discuss the nonreciprocal transport property. 
We calculate the Drude-type nonlinear conductivity, which is closely related to the asymmetric band modulation~\cite{yatsushiro2021microscopic,comment_Hayami}. 
Starting from the second-order Kubo formula, one obtains as 
\begin{align}
\label{eq: Drude}
\sigma_{\eta;\mu\nu}=-\frac{e^3 \tau^2}{2\hbar^3N} \sum_{\bm{k},n}f_{n\bm{k}}\partial_{\eta}\partial_{\mu}\partial_{\nu}\varepsilon_{n\bm{k}}, 
\end{align}
where $\sigma_{\eta;\mu\nu}$ is the Drude-type nonlinear conductivity tensor in $J_{\eta}=\sigma_{\eta;\mu\nu} E_{\mu}E_{\nu}$ for $\eta,\mu,\nu=x,y,z$. 
In Eq.~(\ref{eq: Drude}), $e$, $\tau$, $\hbar$, and $N$ are the electron charge, relaxation time, the reduced Planck constant, and the number of sites, respectively; we set $e=\tau=\hbar=1$ and $N=8\times 2400^2$. 
$\varepsilon_{n\bm{k}}$ and $f_{n\bm{k}}$ are the eigenenergy of the Hamiltonian in Eq.~(\ref{eq: Ham}) and the Fermi distribution function with the band index $n$, respectively (we set the temperature $T=0.01$). 
From the symmetry, nonzero components in the noncoplanar ordered state are given by $\sigma_{x;xx}=-\sigma_{x;yy}=-\sigma_{y;xy}=-\sigma_{y;yx}$; we discuss the bahavior of $\sigma_{x;xx}$. 

Figure~\ref{fig: fig3}(a) shows the contour plot of $\sigma_{x;xx}$ while changing the electron filling per site $n_{\rm e}$ ($n_{\rm e}=2$ corresponds to full filling) and $h$. 
As shown in Fig.~\ref{fig: fig3}(a), $\sigma_{x;xx}$ becomes nonzero in almost all the regions, although their parameter dependences, especially for $n_{\rm e}$ in the small $h$ region, are complicated. 
This might be attributed to the multi-band effect owing to the multi-sublattice system. 
Indeed, $\sigma_{x;xx}$ for low filling monotonically behaves so that the Fermi surface becomes simple [upper panel of Fig.~\ref{fig: fig2}(b)], as shown in the case of $n_{\rm e}=0.18$ in Fig.~\ref{fig: fig3}(b); $|\sigma_{x;xx}|$ developes while increasing $h$. 
It is noted that such a monotonic behavior is unchanged for different hopping parameters, $t_z$, as discussed in Appendix~\ref{sec: Model parameter dependence of nonreciprocal transport}. 
On the other hand, a nonmonotonic behavior including the sign change against $h$ is found in the intermediate $n_{\rm e}$ region, as shown in Fig.~\ref{fig: fig3}(a); we display the result at $n_{\rm e}=0.34$ in Fig.~\ref{fig: fig3}(c), where $\sigma_{x;xx}$ shows several sudden sign reversals in the small $h$ region.

The important hopping process in $\sigma_{x;xx}$ is investigated by expanding $\sigma_{x;xx}$ in the polynomial form of products of $\mathcal{H}^l (\bm{k})$ and the velocity operator, $\bm{v}_{\bm{k}}=\partial \mathcal{H}(\bm{k})/\partial \bm{k}$; $\sigma_{x;xx}=\sum_{ll'l''}C^{ll'l''} \sum_{\bm{k}}{\rm Tr}[v_{x\bm{k}} \mathcal{H}^l (\bm{k})v_{x\bm{k}} \mathcal{H}^{l'} (\bm{k})v_{x\bm{k}} \mathcal{H}^{l''} (\bm{k})]$ where $C^{ll'l''}$ is the model-independent coefficient~\cite{Oiwa_doi:10.7566/JPSJ.91.014701}. 
The contribution to the Drude-type nonlinear conductivity is obtained as the real part after taking the trace. 
Then, the lowest-order contribution to $\sigma_{x;xx}$ is obtained as $h^3 t t^2_z$ for $l=l'=l''=1$, which well corresponds to the model-parameter dependence of the asymmetric band deformation, as discussed above. 
This result indicates that the local SSC over two layers that leads to the asymmetric band modulation plays a key role in inducing the nonreciprocal transport. 
Moreover, by a similar procedure to the asymmetric band modulation, we find that two loops over interlayer triangles (DC'H and CDG) give the lowest-order contribution, as denoted by path 1' and path 2' in Fig.~\ref{fig: fig3}(d). 
Thus, the local SSC $\chi_1$ spanned by DC'H and CDG contributes not only to the asymmetric band modulation but also to nonreciprocal conductivity. 
In particular, one finds a good qualitative correspondence between them for small $n_{\rm e}$ except for their sign in Fig.~\ref{fig: fig3}(b), where the shape of the Fermi surface is simiple~\cite{comment_Hayami}.

\begin{figure}[t!]
\begin{center}
\includegraphics[width=1.0 \hsize ]{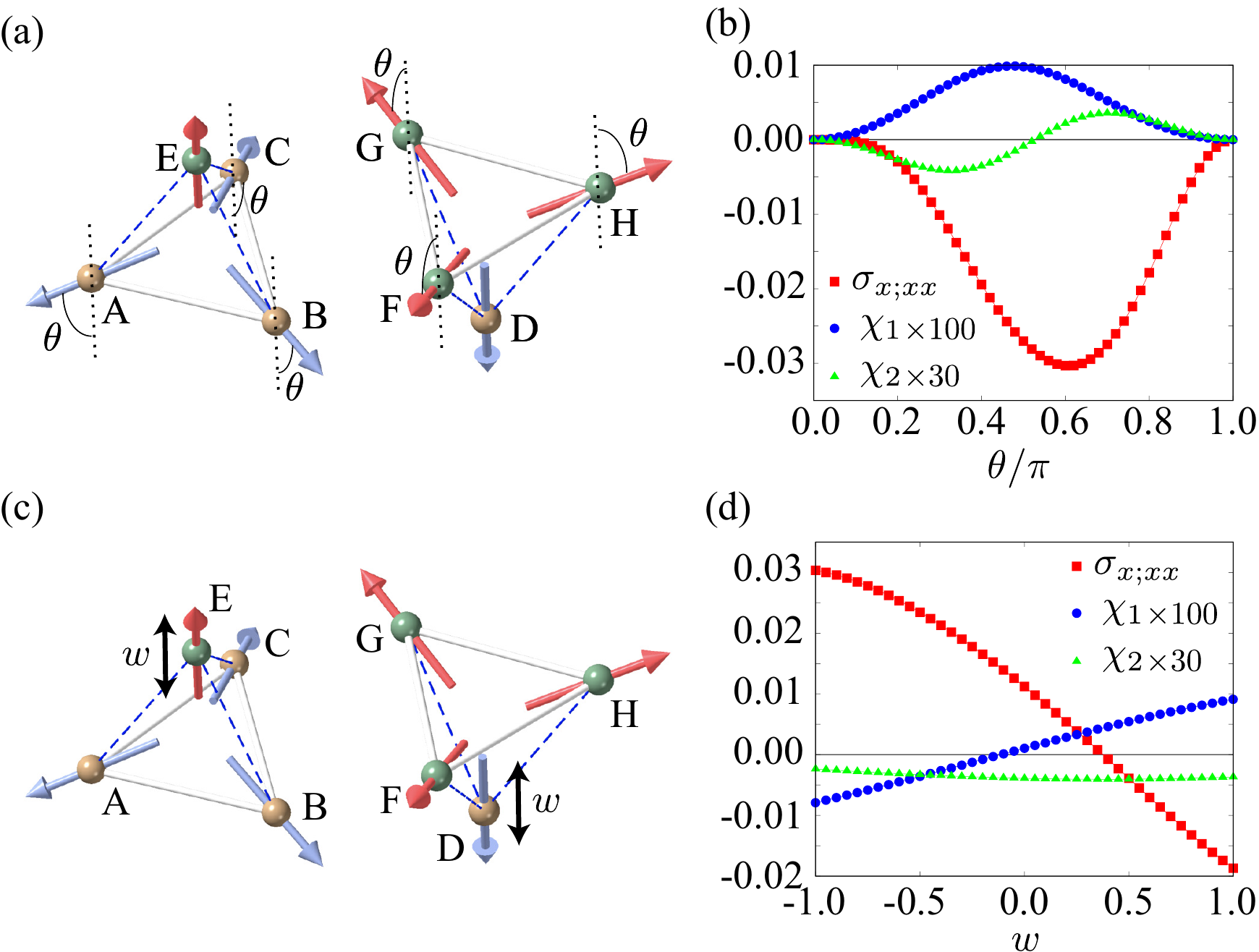} 
\caption{
\label{fig: fig4}
(Color online) 
(a) Schematic picture for the angle modulation $\theta$. 
(b) $\theta$ dependences of $\sigma_{x;xx}$, $\chi_1$, and $\chi_2$ for $n_{\rm e}=0.18$ and $h=2$. 
(c) Schematic picture for the amplitude modulation $w$. 
(d) $w$ dependences of $\sigma_{x;xx}$, $\chi_1$, and $\chi_2$ for $n_{\rm e}=0.18$ and $h=2$. 
}
\end{center}
\end{figure}

To further look into the correspondence between $\sigma_{x;xx}$ and $\chi_1$, we modulate the noncoplanar spin configuration by modifying $\bm{h}_i$ in two ways: One is the angle modulation of $\bm{h}_i$ at sites A--C and F--H and the other is the amplitude modulation of $\bm{h}_i$ at sites D and E. 
For the former, we introduce an angle parameter $\theta$ measured from the negative (positive) $z$ axis for A--C (F--H), as shown in Fig.~\ref{fig: fig4}(a); $\theta=\cos^{-1}(-1/3) \simeq 0.608 \pi$ corresponds to the previous case. 

Figure~\ref{fig: fig4}(b) shows the $\theta$ dependences of $\sigma_{x;xx}$, $\chi_1$, and $\chi_2$ for $n_{\rm e}=0.18$ and $h=2$. 
All the quantities vanish for $\theta=0$ and $\pi$, since the spin configuration reduces to the collinear one. 
$|\sigma_{x;xx}|$ takes the maximum at $\theta \simeq 0.608 \pi$.
The local SSC $\chi_1$ shows the almost symmetric behavior regarding $\theta$ while taking the maxima at $\theta \simeq 0.48 \pi$, whereas $\chi_2$ shows the almost antisymmetric behavior; the sign change of $\chi_2$ is owing to the reversal of all the three spins $\langle s_i^z\rangle$, which leads to the sign reversal of the SSC. 
Thus, $\chi_1$ seems to show a better correspondence to $\sigma_{x;xx}$ compared to $\chi_2$, which is consistent with the microscopic analysis in terms of the model-parameter dependence where the hopping processes on the triangle with $\chi_1$ contributes to the lowest order in $\sigma_{x;xx}$. 
The almost symmetric tendency against $\theta$ is also found in $\sigma_{x;xx}$ at other low and high fillings apart from the half-filling. 
In this region, it has a qualitative correlation to $\chi_1$ rather than $\chi_2$, as presented in Appendix~\ref{sec: Filling dependence of nonreciprocal transport}. 

For the latter modulation, we introduce a weight parameter $w$ for $h_{\rm D}$ and $h_{\rm E}$ like $h_{\rm D} \to w h_{\rm D}$, as shown in Fig.~\ref{fig: fig4}(c), where a negative $w$ presumably causes the sign change of $\chi_1$ while keeping the sign of $\chi_2$. 
We show the behaviors of $\sigma_{x;xx}$, $\chi_1$, and $\chi_2$ at $n_{\rm e}=0.18$ and $h=2$ against $w$ in Fig.~\ref{fig: fig4}(d). 
The sign-reversal behavior is found in $\sigma_{x;xx}$ and $\chi_1$, while it does not occur in $\chi_2$ in Fig.~\ref{fig: fig4}(d), although the $\sigma_{x;xx}$ and $\chi_1$ show the sign change at different $w$, which might arise from the higher-order hopping processes that are relevant to $\chi_2$. 
A similar correspondence between $\sigma_{x;xx}$ and $\chi_1$ is found for the other fillings, as shown in Appendix~\ref{sec: Filling dependence of nonreciprocal transport}. 
On the whole, the results in Figs.~\ref{fig: fig4}(b) and \ref{fig: fig4}(d) indicate that $\sigma_{x;xx}$ has a qualitative correlation to $\chi_1$ when the Fermi surface is relatively simple. 

\section{Summary}
\label{sec: Summary}
To summarize, we have investigated the nonlinear nonreciprocal transport in noncoplanar AFMs. 
By constructing a minimal tight-binding model on the bilayer triangular lattice, we elucidated the essence of nonreciprocal transport driven by the magnetic phase transition. 
We showed that the local SSC degree of freedom becomes a source of the asymmetric band modulation, which results in inducing nonreciprocal transport even without SOC, net SSC, or net magnetization. 
Furthermore, we discussed the important real-space hopping processes in nonreciprocal transport. 
The candidate materials are $X$Nb$_3$S$_6$ ($X=$ Co, Fe, and Ni)~\cite{Tenasini_PhysRevResearch.2.023051,Park_PhysRevMaterials.6.024201} and Ru$_2$Mn$X'$ ($X'=$ Si and Sb)~\cite{Simon_PhysRevMaterials.4.084408}, where similar noncoplanar spin textures have been suggested.

\appendix

\section{Model parameter dependence of nonreciprocal transport}
\label{sec: Model parameter dependence of nonreciprocal transport}

\begin{figure}[t!]
\begin{center}
\includegraphics[width=0.5 \hsize ]{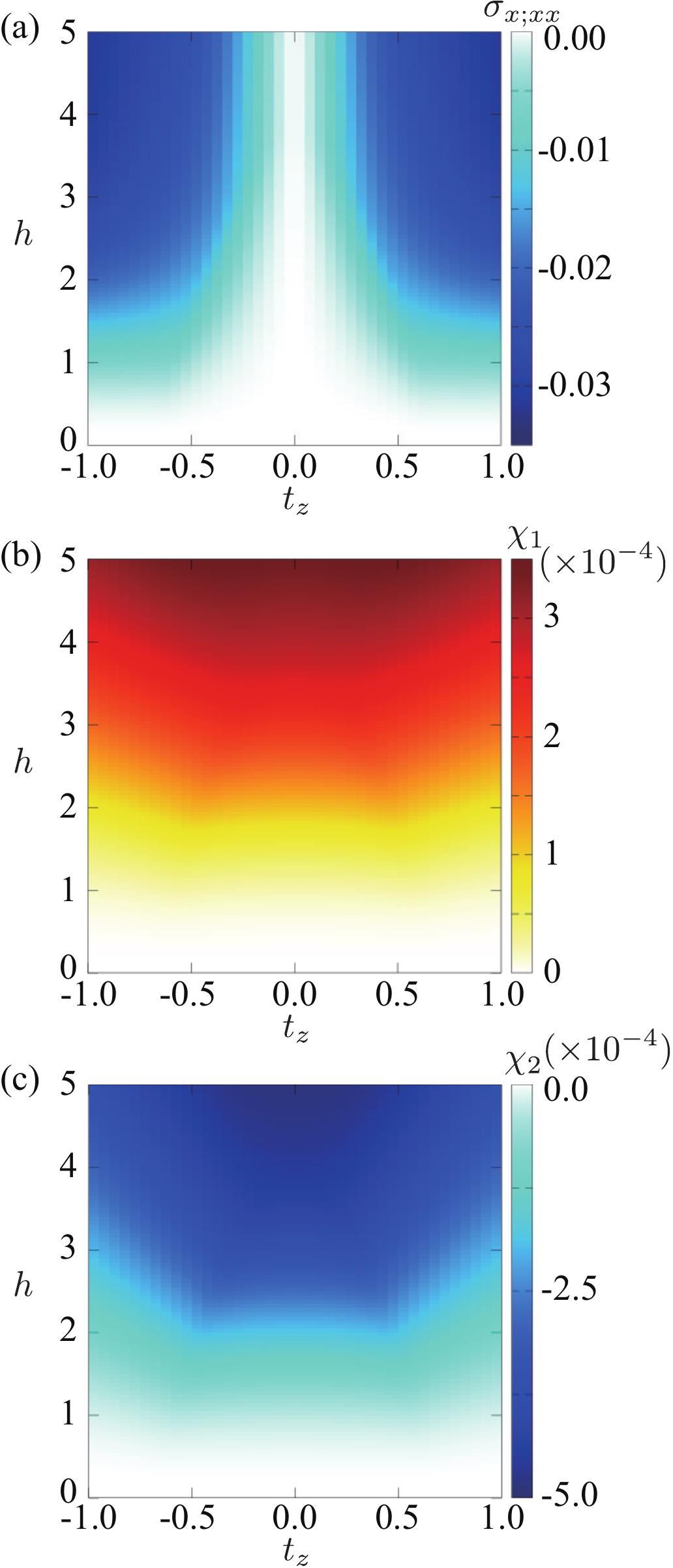} 
\caption{
\label{fig: app1}
(Color online) 
Contour plot of (a) $\sigma_{x;xx}$, (b) $\chi_1$, and (c) $\chi_2$ in the plane of $t_z$ and $h$ for low filling $n_{\rm e}=0.18$. 
}
\end{center}
\end{figure}

We present the behaviors of $\sigma_{x;xx}$, $\chi_1$, and $\chi_2$ as functions of $t_z$ and $h$ for low filling $n_{\rm e}=0.18$ in Figs.~\ref{fig: app1}(a), \ref{fig: app1}(b), and \ref{fig: app1}(c), respectively. 
As shown in Fig.~\ref{fig: app1}(a), $|\sigma_{x;xx}|$ increases while increasing $h$ independent of $t_z$. 
It is noted that $\sigma_{x;xx}$ vanishes for $t_z=0$, which is consistent with the model parameter analysis in Sect.~\ref{sec: Nonreciprocal transport} in the main text. 
In Figs.~\ref{fig: app1}(b) and \ref{fig: app1}(c), both $\chi_1$ and $\chi_2$ also show the monotonic behaviors against $h$. 
Thus, $h$ dependence of $\sigma_{x;xx}$, $\chi_1$, and $\chi_2$ appears to be similar for low filling so that the Fermi surface becomes simple even when changing $t_z$ except for $t_z=0$.

\section{Filling dependence of nonreciprocal transport}
\label{sec: Filling dependence of nonreciprocal transport}

\begin{figure}[t!]
\begin{center}
\includegraphics[width=1.0 \hsize ]{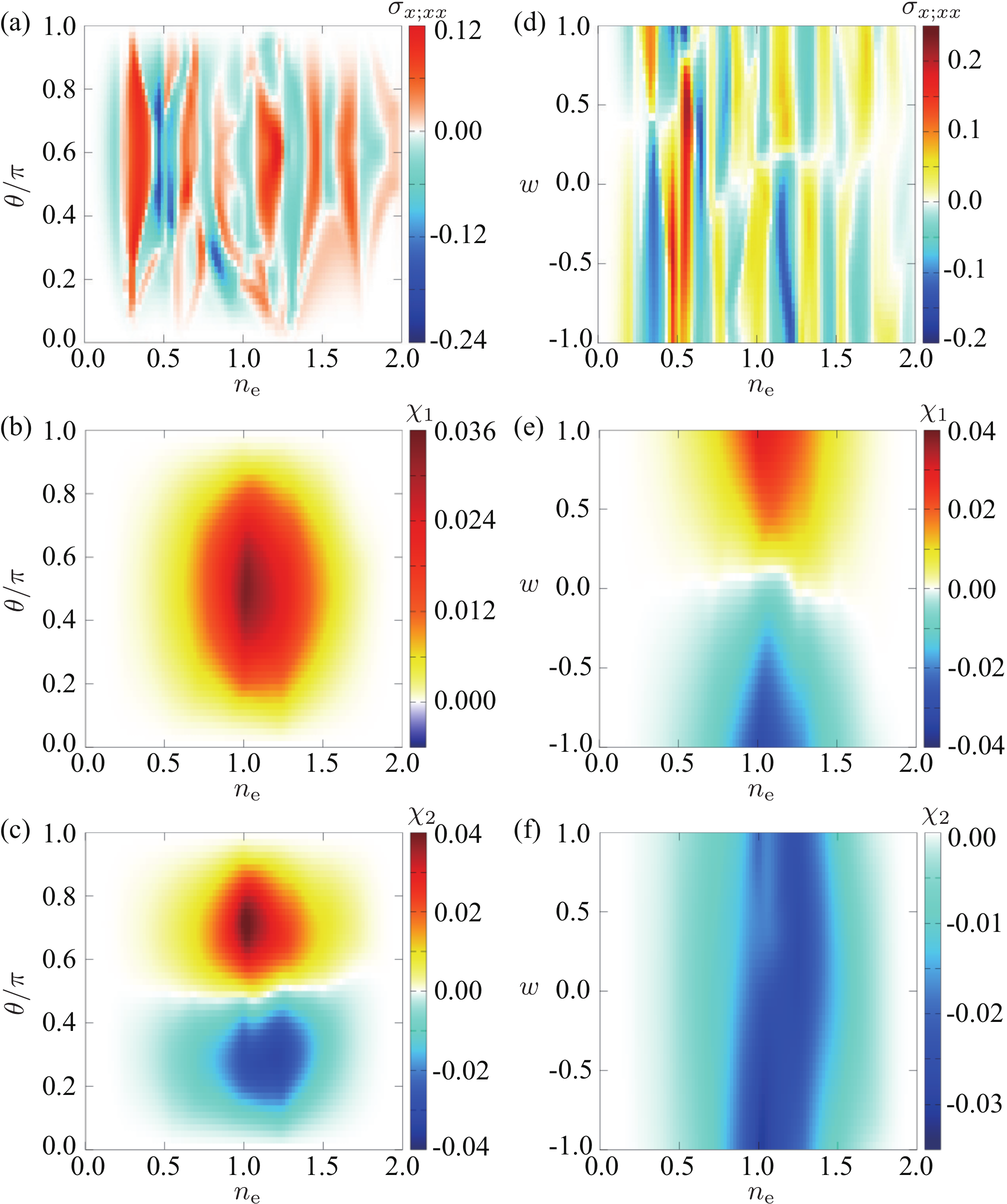} 
\caption{
\label{fig: app2}
(Color online) 
Contour plot of (a,d) $\sigma_{x;xx}$, (b,e) $\chi_1$, and (c,f) $\chi_2$ in the plane of $n_{\rm e}$ and (a-c) $\theta$ [(d-f) $w$], which corresponds to the situation in Fig.~\ref{fig: fig4}(a) [Fig.~\ref{fig: fig4}(c)]. 
}
\end{center}
\end{figure}

We show the filling dependence of $\sigma_{x;xx}$, $\chi_1$, and $\chi_2$ while changing $\theta$ and $w$, which are introduced in Figs.~\ref{fig: fig4}(a) and \ref{fig: fig4}(c) in the main text, respectively, in Figs.~\ref{fig: app2}(a)--\ref{fig: app2}(f). 
Figure~\ref{fig: app2}(a) shows $\sigma_{x;xx}$ as functions of $n_{\rm e}$ and $\theta$, where the data at $n_{\rm e}=0.18$ corresponds to those in Fig.~\ref{fig: fig4}(b). 
The data indicate that $|\sigma_{x;xx}|$ tends to become larger for $\theta \simeq 0.5 \pi$ and be a symmetric in terms of $\theta$, especially for low filling. 
Such behavior has a correspondence to $\chi_1$ rather than $\chi_2$, as shown in Figs.~\ref{fig: app2}(b) and \ref{fig: app2}(c).

Similarly, the $w$ dependence of $\sigma_{x;xx}$ in Fig.~\ref{fig: app2}(d) is characterized by the antisymmetric behavior against $w$ independent of $n_{\rm e}$; $\sigma_{x;xx}$ shows the sign change while changing $w$ at fixed $n_{\rm e}$. 
As discussed in Sect.~\ref{sec: Nonreciprocal transport} in the main text, such a tendency is similar to that of $\chi_1$ instead of $\chi_2$, the former of which shows the sign change against $w$.

\begin{acknowledgments}
S.H. thanks S. Seki for useful discussions. 
This research was supported by JSPS KAKENHI Grants Numbers JP19K03752, JP19H01834, JP21H01037, JP22H04468, JP22H00101, JP22H01183, and by JST PRESTO (JPMJPR20L8).
Parts of the numerical calculations were performed in the supercomputing systems in ISSP, the University of Tokyo.
\end{acknowledgments}

\bibliographystyle{JPSJ}
\bibliography{ref}

\end{document}